\documentstyle[prl,aps,twocolumn,psfig,floats]{revtex}
\input{epsf}
\begin{document}
\draft
\twocolumn[\hsize\textwidth\columnwidth\hsize\csname@twocolumnfalse\endcsname

\title{The Phase Diagram of Disordered Vortices from London 
       Langevin Simulations}
\author{Anne van Otterlo$^{\dagger}$, Richard T. Scalettar
  and Gergely T. Zim\'{a}nyi}
\address{Physics Department, University of California, Davis, CA 95616, USA}
\date{February 18, 1998}
\maketitle

\begin{abstract}
  We study the phase diagram of vortex matter in disordered type-II
  superconductors. We performed numerical simulations in the London
  Langevin approximation, using a new realistic representation of the
  disorder.  At low magnetic fields we find a disentangled and
  dislocation free Bragg-glass regime.  Increasing the field
  introduces disorder-driven entanglement in a discontinuous manner,
  leading to a vortex-glass phase, which subsequently melts into the
  vortex liquid.  The obtained phase boundaries are in quantitative
  agreement with the experimental data.
\end{abstract}
]

\begin{narrowtext}

Fluctuations in the High Temperature Superconductors (HTSC) play a
dominant role in vortex phenomenology due to the high transition
temperatures $T_{C}\sim$100K and large anisotropies
$\epsilon^{-1}\sim$5-500.  As a result, in clean systems the Abrikosov
lattice melts into an entangled vortex liquid with increasing
temperature.  This transition was predicted on the basis of Lindemann
type calculations~\cite{nelson88,bfglv94} and is by now firmly
established by London~\cite{henrik97}, 3D XY~\cite{tachiki97}, and
Ginzburg-Landau~\cite{ginlan95} simulations.  Experimentally, the
latent heat found in YBCO~\cite{schilling96} and the jump in the
magnetization in BSSCO~\cite{delacruz94,zeldov95} and
YBCO~\cite{bonn96,welp96} provide thermodynamic evidence for a first
order melting transition at low fields.

Although disorder was first argued to eliminate the long range order
of the Abrikosov lattice altogether~\cite{larkov73}, more recent
theories propose that quasi-long-range translational order is still
preserved at low magnetic fields in a dislocation free ``Bragg-glass''
phase~\cite{natter90,gialed94}.  Increasing the field effectively
increases the disorder and as the pinning energies overcome the
elastic energies, vortices entangle~\cite{ertas96}
forming a vortex-glass~\cite{fisher89}.  Experimental support for this
scenario is the sharp enhancement of the critical current from
magnetization measurements~\cite{hardy94,klein94,zeldov96,groot97},
the rapid destruction of the Bragg peaks in neutron
scattering~\cite{forgan93}, and the pronounced dips in the I-V
curves~\cite{safar95}.  A numerical study found evidence for pairs of
point-like disclination defects driving the Bragg-glass to vortex
glass transition in layered systems with strong pins~\cite{ryu96}.
Transport measurements~\cite{safar93,kwok94} indicate that the first
order melting transition from the Bragg-glass to the vortex liquid at
low fields is replaced by a continuous melting transition from the
vortex glass to the vortex liquid at higher fields.  This basic
feature is reproduced by numerical
simulations~\cite{jagla96,wilkin97}.  Whether the vortex glass and
vortex liquid are separated by a phase transition or a crossover is
presently under discussion and seems to depend on the screening
length~\cite{young97}.

In this Letter we report continuum Langevin simulations for disordered
vortex systems in the London approximation, which is valid outside the
3D XY critical region and for fields $B\lesssim$0.2$H_{c2}(T)$.  This
approach is more realistic than lattice simulations due to the absence
of intrinsic pinning, yet it is capable of reaching satisfactory
system sizes.  Furthermore, by using a coarse grained pinning
landscape, we are able to reach the weak collective pinning regime.
We measure the in-plane structure factor $S({\bf Q})$, the vortex
wandering along the field $B(z)$, as well as the time overlap
correlator $C(t)$ (see below for the definitions of these quantities).
The central result of our paper is the determination of the phase
diagram for realistic YBCO parameters (Fig.~1).

\begin{figure}[hbt] \unitlength1cm \begin{picture}(9.0,7.0)
  \put(-0.5,0.0)
  {\psfig{figure=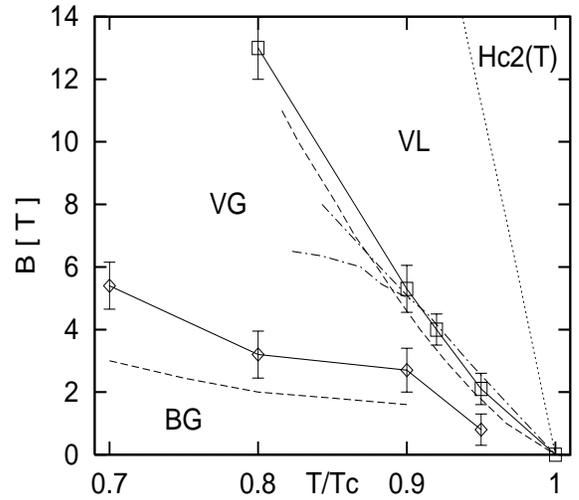,height=7.0cm,width=9.0cm,angle=0}}
\end{picture}
\caption{The $B$--$T$ phase diagram for YBCO parameters,
  consisting of a Bragg Glass, Vortex Glass, and Vortex Liquid phases.
  For comparison, experimental data is shown as dashed
  (Ref.~\protect\cite{groot97}) and dash-dotted
  (Ref.~\protect\cite{safar95}) lines.}
\end{figure}

We start by outlining our numerical approach.  The magnetic field is
taken to be in the $c$-direction, perpendicular to the
CuO$_{2}$-layers, indexed by $z$.  The Langevin equation describing
the overdamped motion of vortices reads
\begin{equation}
  \eta\frac{\partial{\bf R}_{\mu}(z,t)}{\partial t}=
  {\bf \zeta}_{\mu}(z,t)+{\bf F}_{L} -\frac{\delta H[\{{\bf
      R}_{\nu}(z,t)\}]} {\delta {\bf R}_{\mu}(z,t)}\;,
\end{equation}
were $\mu$ labels the vortices with coordinates ${\bf R}$.  The
damping force on the left hand side is characterized by the
coefficient $\eta$ that is related to the Bardeen-Stephen flux-flow
resistivity by $\rho_{\rm BS}= B\Phi_{0}/(c^{2}\eta)$, where
$\Phi_{0}= hc/2e$ is the flux quantum.  The Langevin force ${\bf
\zeta}$ is correlated as $\langle\zeta^{\alpha}_{\mu}(z,t)
\zeta^{\beta}_{\nu}(z',t')\rangle= 2\eta T\delta^{\alpha\beta}
\delta_{\mu\nu} \delta(z-z')\delta(t-t')$, with $\alpha$, $\beta$=x,y.
The Lorentz force from a bias current density ${\bf J}$ is ${\bf
F}_{L}= \Phi_{0}{\bf J}\times
\hat{\bf z}/c$.  The Hamiltonian $H$ is constructed on the basis of
London theory.  Its derivative decomposes into three forces.  The
pairwise interaction force is the periodic extension of the in-plane
London interaction, in Fourier space $V(k)= 2\pi\epsilon_{0}
\exp(-k^{2}\xi^{2})/ (\lambda^{-2}+k^{2})$, with line energy
$\epsilon_{0}= \Phi^{2}_{0}/ (8\pi\lambda)^{2}$, London penetration
depth $\lambda$, and coherence length $\xi$.  We choose
$\kappa$=$\lambda/\xi$=100. The bending forces are governed by the
single vortex elastic constant $c_{44}=\epsilon^{2} \epsilon_{0}
\log(\kappa)$, with anisotropy $\epsilon$=1/5.  Finally, the
dimensionless pinning force on vortex $\mu$ in layer $z$ for $\delta
T_{C}$ disorder is given by
\begin{equation}
  {\bf F}^{P}_{\mu,z}=-\int d^{2}R' {\bf \nabla}_{{\bf R}_{\mu}(z)}
  p({\bf R}_{\mu}(z)-{\bf R}') u({\bf
    R}',z)\;,
\end{equation} 
where the microscopic disorder potential is delta-correlated, $\langle
u({\bf R},z)u({\bf R}',z')\rangle= \gamma\delta({\bf R}-{\bf
R}')\delta_{z,z'}$, with strength $\gamma$.  With our conventions, the
parameter $\gamma$ equals the parameter $\delta_{\alpha}$ in Chapter
IIIC of Ref.~\cite{bfglv94}.  The convolution with the long-range
vortex form factor $p(R)=2\xi^{2}/(R^{2}+2\xi^{2})$ generates a smooth
pinning landscape.

\begin{figure}[hbt] \unitlength1cm
\begin{picture}(9.0,4.7)
  \put(0.,-0.8)
  {\psfig{figure=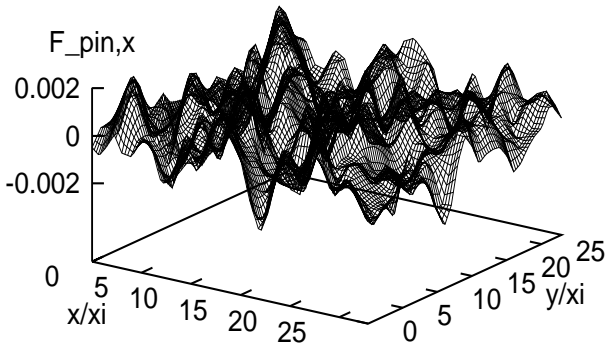,height=7.0cm,width=9.0cm,angle=0}}
\end{picture}
\caption{A part of a typical pinning landscape (``pinscape''). Shown is
  the pinning force in the $x$-direction.}
\end{figure}

The $z$ coordinate is discretized with the natural layer spacing
$d\approx 12$\AA.  The total number of layers in the $z$-direction is
denoted by $L_{z}$.  We employ {\em open} boundary conditions in this
direction, which allows topological defects to enter the sample from
the top and bottom.  The in-plane coordinates ${\bf R}_{\mu}$ satisfy
periodic boundary conditions and lie in a parallelogram spanned by the
unit vectors $\hat{e}_{a}=\hat{e}_{x}$ and $\hat{e}_{b}=
(\hat{e}_{x}+\sqrt{3}\hat{e}_{y})/2$ with area given by the number of
vortices $N$ times $\Phi_{0}/B$.  The unit of length is taken to be
the coherence length $\xi(T)$, with $\xi(0)$=17\AA, so that
$H_{c2}(0)$=120T; the average inter-vortex distance is $a_{\Phi}$.
The natural unit of time is $t_{0}=
\xi^{2}\eta/\epsilon_{0}$ and the time $t$ is discretized with
timestep $\delta t_{0}$; we take $\delta$=0.05.  We mention two
technical aspects of importance.  Considerable gain is reached in
computational speed by generating and storing the pinning forces (the
``pinscape'', see Fig.~2) on a regular array of spacing $\xi$.  The
pinning force at a specific vortex location is then constructed by
interpolating between the four closest grid-points.  Second, in order
to equilibrate disordered systems, we employed simulated annealing,
cooling from $\sim 2.5 T$ to $T$ in 10 {\%} decrements.  Most of our
simulations were done for $7\times 7$ vortices in 50 layers, using
100,000 sweeps for equilibration and 100,000 for measurement.  With
the exceptions below we found only limited fluctuations between
different disorder configurations, due to the good self-averaging of
the $\sim 2500$ vortex elements.  Therefore, typically we averaged
over only a few disorder realizations.  Finally the disorder strength
was chosen to yield reasonable low temperature values for the critical
current in terms of the depairing current $j_{0}$: $j_{c}/j_{0} \sim
0.005$.

We now proceed to investigate the transitions between the three
proposed phases, the dislocation free Bragg Glass (BG), the dislocated
Vortex Glass (VG), and the Vortex Liquid (VL).

\begin{figure}[hbt] \unitlength1cm
\begin{picture}(9.0,7.0)
  \put(-0.2,0.0)
  {\psfig{figure=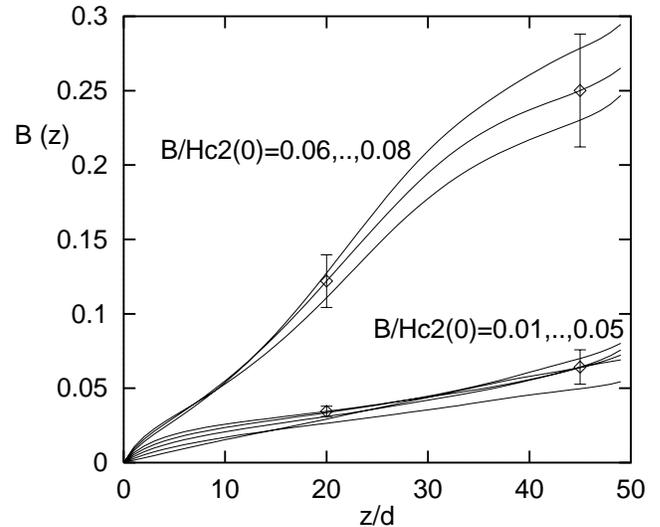,height=7.0cm,width=9.0cm,angle=0}}
\end{picture}
\caption{The vortex wandering $B(z)$ in the z-direction for different
  magnetic fields at $T/T_{c}=0.7$, $\gamma$=0.1, and $\epsilon$=0.2
  shows the Bragg to Vortex Glass transition.}
\end{figure}

{\bf The BG-VG transition.} In the dislocation-free Bragg Glass
vortices fluctuate in the self-consistent cage of their
neighbours~\cite{ertas96}.  Increasing the field brings vortices
closer together, and increases the relative strength of the disorder.
When the single vortex wandering reaches the Lindemann limit, vortices
entangle and dislocations enter the sample~\cite{ertas96,giam2}.  Our
measured vortex line wandering, $B(z)= <\![{\bf R}_{\mu}(z,t)- {\bf
R}_{\mu}(0,t)]^{2}>\!/a^2_{\Phi}$, is displayed in Fig.~3 for a
typical field sweep.  The curves for the five lowest fields group
together and belong to the Bragg glass, whereas the curves for the
higher fields are in the vortex glass phase.  Our low field data in
the BG in Fig.~3 grow slowly with $z$, and are consistent with a power
law with exponent less than unity.  The presently attainable system
sizes are, however, too small to see the asymptotic regime, where
$B(z)\sim ln(z)$~\cite{gialed94}.  Increasing the field across the
transition enhances $B(z)$ abruptly and instead of bending down, the
curves now bend upward until boundary effects become important around
$z\sim L_{z}$.

The vortex wandering in the VG phase reaches the Lindemann number
$c_{L}$ at the entanglement length in the $z$-direction $L_{E}$.  From
$B(L_{E})\equiv c^{2}_{L}\lesssim$ 0.1, we deduce that just above the
transition, the entanglement length is of the order of 15 lattice
spacings in the $z$-direction for the parameters of Fig.~3.  Thus, our
system size in the $z$-direction (50) is large enough to see
entanglement, as evidenced by the snapshot just inside the VG phase in
Fig.~4b, which shows two entangled ``ring-exchanges'', or
screw-dislocation loops, involving 2 and 9 vortices.  On the other
hand, Fig.~4a, taken on the BG side of the transition shows a perfect
disentangled lattice.

\begin{figure}[hbt] \unitlength1cm \begin{picture}(9.5,10.7)
  \put(-0.6,5.6)
  {\psfig{figure=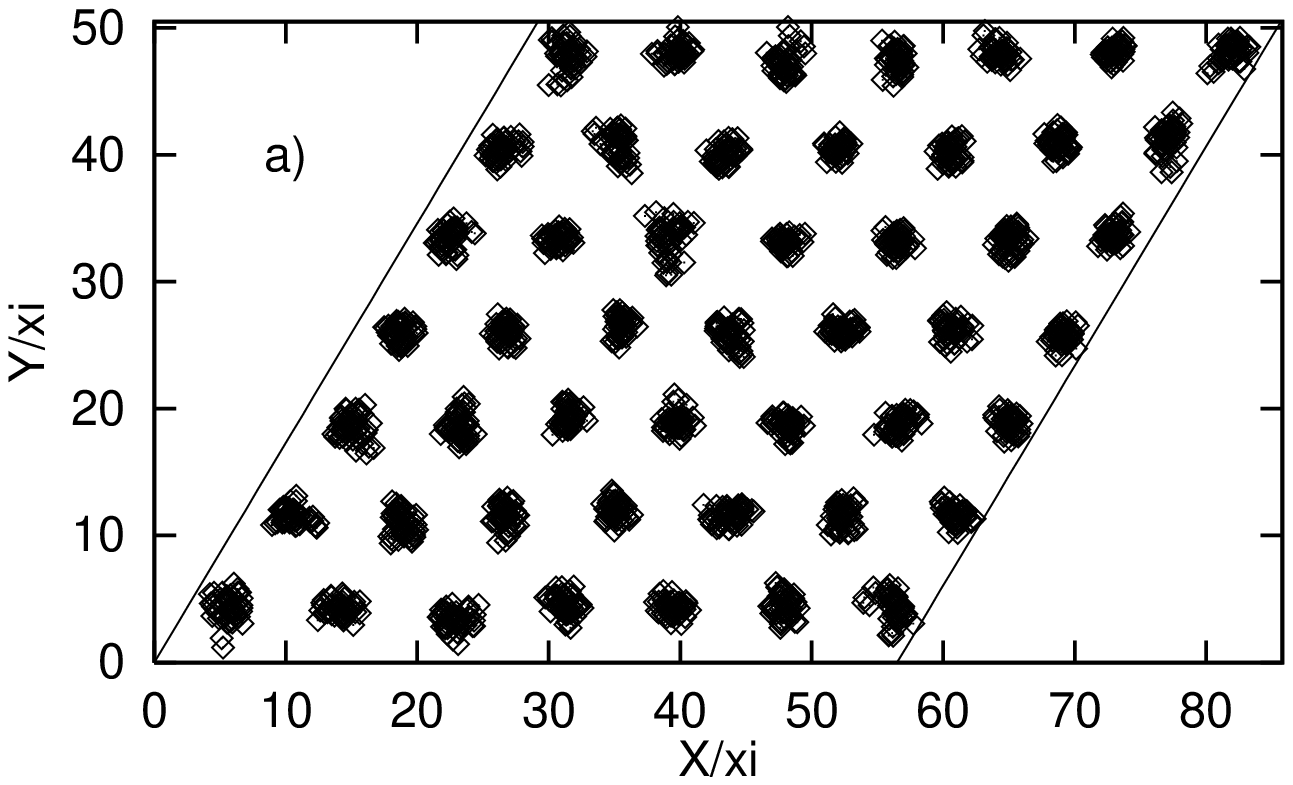,height=5.2cm,width=9.0cm,angle=0}}
  \put(-0.6,0.2)
  {\psfig{figure=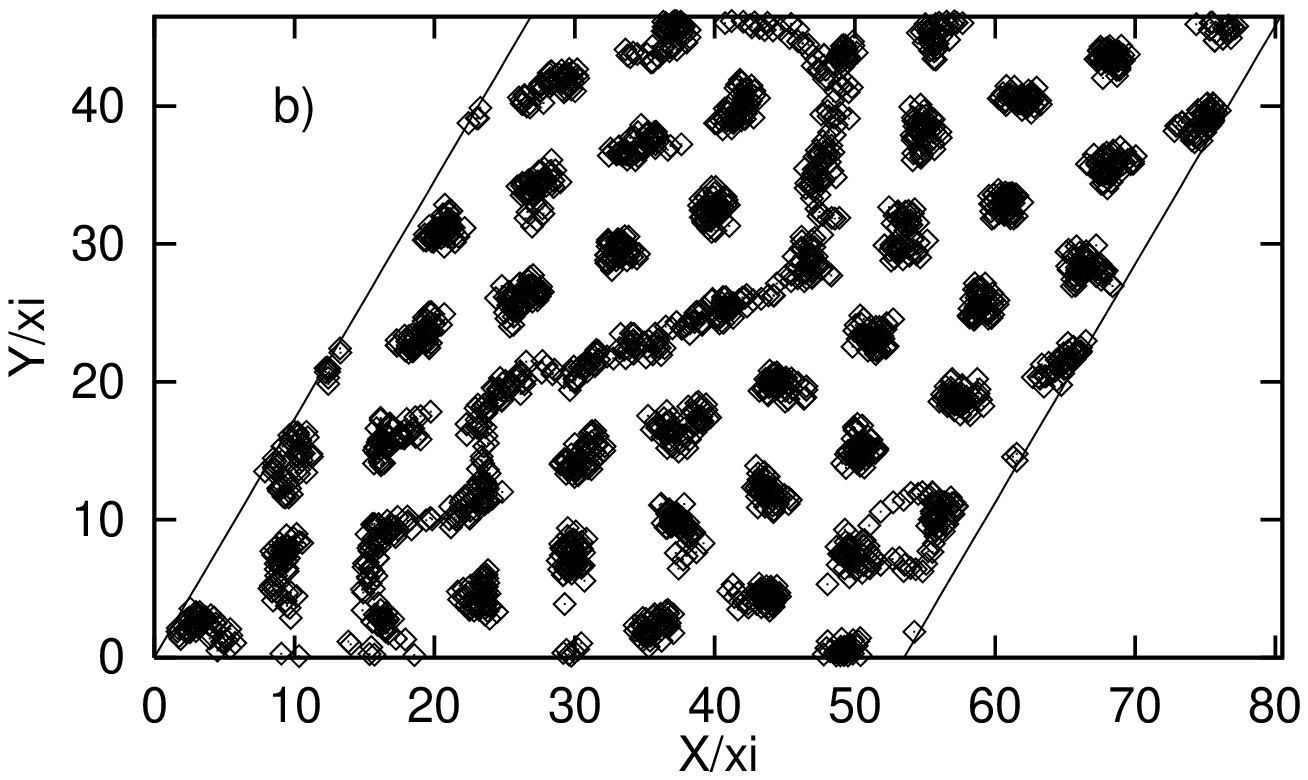,height=5.2cm,width=9.0cm,angle=0}}
\end{picture}
\caption{Snapshots of a typical configuration projected in the
  $xy$-plane in the a) the Bragg glass phase at $B$=6T and b) the
  Vortex Glass phase at $B$=7.2T.  Parameters as for
  Fig.~3.}
\end{figure}

The average height of the six principle structure factor peaks in
$S({\bf Q})= <\!(1/N)\sum_{\mu,\nu}\exp(i{\bf Q}\cdot[{\bf
R}_{\mu}(z,t)-{\bf R}_{\nu}(z,t)])\!>$ is also indicative of the BG-VG
transition.  This is shown in Fig.~5, together with the wandering
across the sample $B(L_{z})$.  In the BG phase, the structure factor
first increases with field, due to the growing interaction strength.
Upon crossing the BG-VG transition, the structure factor peak height
drops at the same point where the wandering increases dramatically.
Frozen-in short range order still persists in the VG phase, however,
as is clear from Fig.~4b.

\begin{figure}[hbt] \unitlength1cm
\begin{picture}(9.0,6.0)
  \put(-0.5,0.3)
  {\psfig{figure=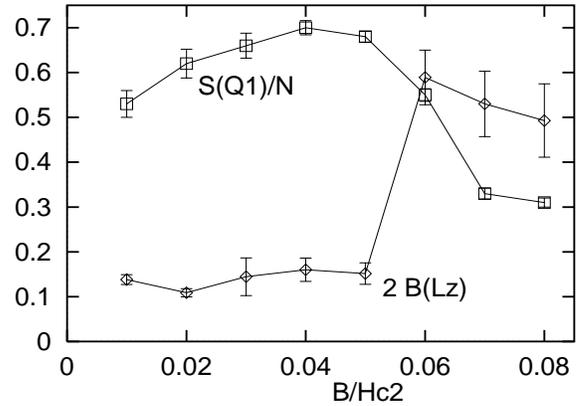,height=5.5cm,width=9.0cm,angle=0}}
\end{picture}
\caption{The magnetic field dependence at $T/T_{C}$=0.7 of the
  averaged amplitude of the first 6 primary structure factor peaks
  (boxes), normalized to the number of vortices $N$=49, and the
  wandering across the sample $B(L_{z})$ (diamonds).}
\end{figure}

For different disorder realizations, the jump in $B(L_{z})$ and the
corresponding drop of the structure factor peaks, occur at different
magnetic fields.  As a result, a disorder average of the quantities we
measure shows only a smooth field dependence.  Although the sharp jump
in the wandering for a single disorder realization is consistent with
a first order transition, we speculate that in the thermodynamic limit
only a broad feature survives, which mimics a continuous transition or
a cross-over.

{\bf VG-VL transition.} This transition is defined by a profound
change in the dynamical behavior of the system: the vortex glass can
be seen as a frozen liquid.  This is most clearly seen in the behavior
of the time overlap correlator $C(t) =\exp(-<[{\bf R}_{\mu}(z,t)-{\bf
  R}_{\mu}(z,0)]^{2}> /l^2)$, as plotted for $l$=0.2$a_{\Phi}$ in the
inset of Fig.~6.  For times $t\gg t_{0}$ it decays exponentially with
a ``plastic'' relaxation time $t_{pl}$, which is plotted as a function
of temperature in Fig.~6.  Upon cooling from the VL to the VG, the
time $t_{pl}$ rapidly increases, signalling that vortices get trapped
in the almost static cages of the surrounding vortices.  The plastic
time does not appear to be infinite in the VG, however.  It saturates
for temperatures below the VG-VL line, at a value around
$t_{pl}\sim$ 2000$t_{0}$.  Whether this is a finite size effect, or
hints to the absence of a genuine phase transition with universal
exponents, will be the subject of a future finite size
study~\cite{future}.

\begin{figure}[hbt] \unitlength1cm \begin{picture}(9.5,7.5)
  \put(-1.0,0.0)
  {\psfig{figure=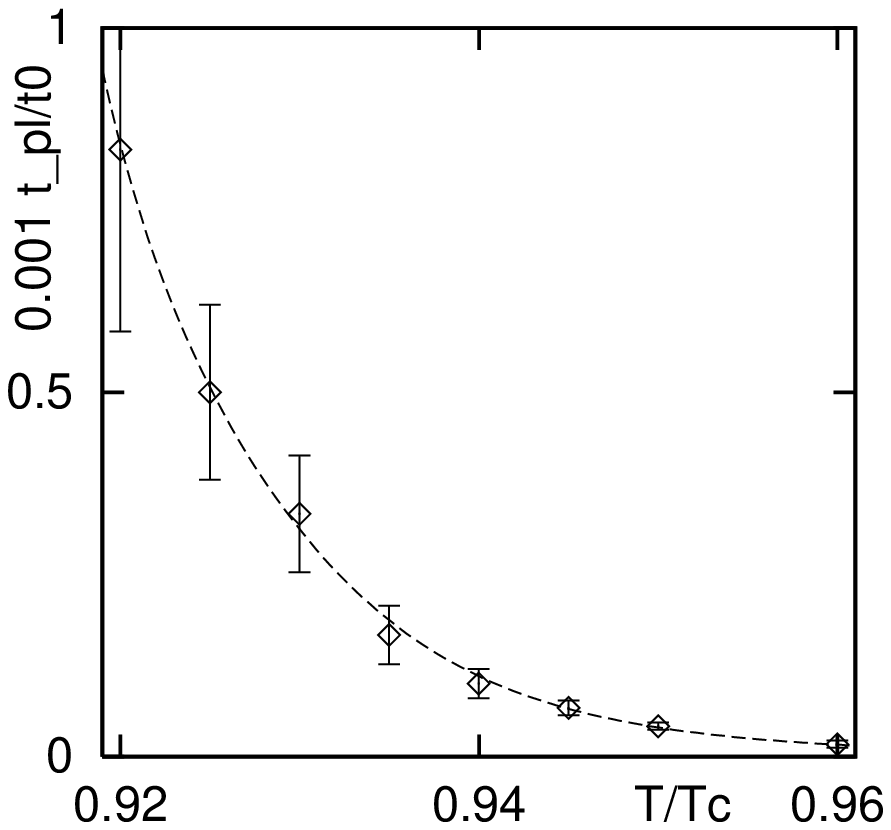,height=7.5cm,width=9.5cm,angle=0}}
  \put(1.5,1.9)
  {\psfig{figure=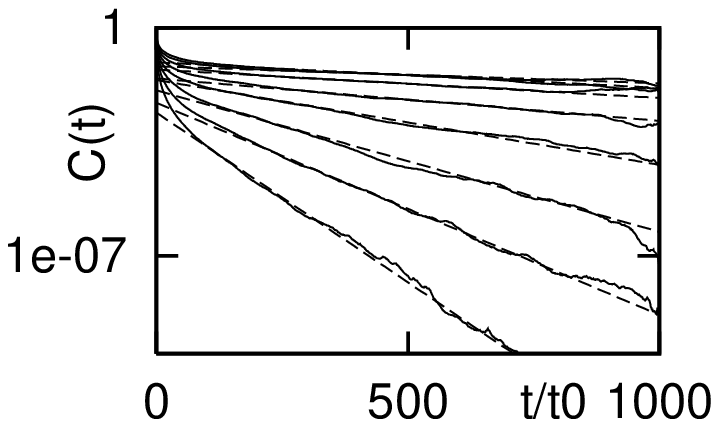,height=5.3cm,width=6.5cm,angle=0}}
\end{picture}
\caption{The relaxation time $t_{pl}/(10^{3}t_{0}$) as a function of
  temperature from an exponential fit to the correlator $C(t)$
  illustrates the VG-VL transition. Here $B$=4T and the dashed line is
  a guide to the eye. The inset shows the correlation function $C(t)$
  as a function of time for $T/T_{c}$= 0.91, .., 0.95 (step 0.005,
  full lines, from top to bottom) and the fits (dashed lines).}
\end{figure}

As a function of temperature and field, the six primary structure
factor peaks decrease, but in the vortex-glass phase with frozen short
range order, they remain finite and visible in our finite system and
simulation time.  As the vortex glass melts into the liquid, the
primary peaks dissolve into a ringlike structure.  Transport
measurements can also identify the VG-VL transition.  Fisher et
al.~\cite{fisher89} suggested that the flux flow resistivity at high
temperature gives way to non-ohmic exponential creep below the glass
transition temperature, a prediction which was confirmed
experimentally~\cite{koch}, see, however, also Ref.~\cite{lopez98}.
Our numerical results on the IV curves will be presented
elsewhere~\cite{future}.

The above diagnostics allow us to construct the phase diagram.  We
identified the predicted dislocation free Bragg Glass, entangled
Vortex Glass, and Vortex Liquid phases in the $B$--$T$ plane, as
summarized in Fig.~1.  We find that the BG-VG phase boundary is rather
sensitive to the disorder, whereas the VG-VL transition is not.  In
fact, even from disorder realization to realization, the location of
the BG-VG boundary shows detectable fluctuations.  This sensitivity is
a well known experimental fact: Safar et al.~\cite{safar95} report
transition fields from $3T$ to $10T$ in samples with minute
macroscopic differences, whereas the data of Ref.~\cite{klein94} are
at even lower fields. Experimentally, the BG-VG transition appears as
a sharp jump in the magnetization only in BSSCO, while in YBCO and
Thallium based materials the transition is rather smooth. The reported
numerical fluctuations seem to be related to this smearing of the
BG-VG transition in the latter compounds.  In Fig.~1 we also display
the experimental data of Deligiannis et al.~\cite{groot97} and Safar
et al.~\cite{safar95}.  We chose a disorder value to make the
transition field fall between the two sets of data. The correspondence
between the simulations and experimental data is remarkable,
especially for the VG-VL transition.  At the highest temperatures
$T/T_c \sim 0.95$, the BG-VG phase boundary bends downward.  The
resolution of our data is insufficient at the moment to decide,
whether the BG-VG line ends somewhere on the melting line, as the
BSSCO~\cite{zeldov95} and some YBCO data~\cite{safar95} indicates, or
whether there is a sliver of the VG phase intruding all the way down
to zero fields, preempting a direct BG-VL transition, as the
Thallium~\cite{hardy94} and other YBCO data~\cite{klein94,groot97}
suggests.

We thank G. Blatter, T. Giamarchi, P. Kes, and V. Vinokur for
discussion.  We gratefully acknowledge the hospitality of the ETH in
Z\"{u}rich (AvO) and of the Argonne National Laboratories (GTZ).  This
research was supported by NSF grant DMR 95-28535 and by the Dutch
Foundation for Fundamental Research on Matter (FOM).

\vspace{10mm}

$^{\dagger}$ Present address: Instituut-Lorentz for Theoretical
Physics, P.O.B. 9506, 2300 RA Leiden, The Netherlands.

\end{narrowtext}

\vspace{-21mm}

\begin{references}
\bibitem{nelson88} D.~R. Nelson, Phys. Rev. Lett. {\bf 60}, 1973 (1988).
\bibitem{bfglv94} G. Blatter et al.,
  Rev. Mod. Phys. {\bf 66}, 1125 (1994).
\bibitem{henrik97} H. Nordborg and G. Blatter,
  Phys. Rev. Lett. {\bf 79}, 1925 (1997).
\bibitem{tachiki97} X. Hu, S. Miyashita, and M. Tachiki,
  Phys. Rev. Lett. {\bf 79}, 3498 (1997); S. Ryu and D. Stroud,
  Phys. Rev. Lett. {\bf 78}, 4629 (1997).
\bibitem{ginlan95} R. Sasik and D. Stroud,
  Phys. Rev. Lett. {\bf 75}, 2582 (1995); J. Hu and A. MacDonald,
  Phys. Rev. B {\bf 56}, 2788 (1997).
\bibitem{schilling96}   A. Schilling et al., Nature {\bf 382}, 791 (1996).
\bibitem{delacruz94} H. Pastoriza et al.,
  Phys. Rev. Lett. {\bf 72}, 2951 (1994).
\bibitem{zeldov95} E. Zeldov et al., Nature {\bf 375}, 373 (1995).
\bibitem{bonn96} R. Liang, D.~A. Bonn, and W.~N. Hardy,
  Phys. Rev. Lett. {\bf 76}, 835 (1996).
\bibitem{welp96} U. Welp et al., Phys. Rev. Lett. {\bf 76}, 4809 (1996).
\bibitem{larkov73} A.~I. Larkin and Yu.~N. Ovchinnikov,
  Zh. Eksp. Teor. Fiz. {\bf 65}, 1704 (1973)
  [Sov. Phys. JETP {\bf 38}, 854 (1974)].
\bibitem{natter90} T. Natterman, Phys. Rev. Lett. {\bf 64}, 2454 (1990).
\bibitem{gialed94} T. Giamarchi and P. Le Doussal,
  Phys. Rev. Lett. {\bf 72}, 1530 (1994);
  Phys. Rev. B {\bf 52}, 1242 (1995).
\bibitem{ertas96} D. Ertas and D.~R. Nelson,
  Physica C {\bf 272}, 79 (1996).
\bibitem{fisher89} M.~P.~A. Fisher,
  Phys. Rev. Lett. {\bf 62}, 1415 (1989);
  D.~S. Fisher, M.~P.~A. Fisher, and D.~A. Huse,
  Phys. Rev. B {\bf 43}, 130 (1991).
\bibitem{hardy94} V. Hardy et al., Physica C {\bf 232}, 347 (1994).
\bibitem{klein94} L. Klein et al., Phys. Rev. B {\bf 49}, 4403 (1994).
\bibitem{zeldov96} B. Khaykovich et al.,
  Phys. Rev. Lett. {\bf 76}, 2555 (1996).
\bibitem{groot97} K. Deligiannis et al.,
  Phys. Rev. Lett. {\bf 79}, 2121 (1997).
\bibitem{forgan93} R. Cubitt et al., Nature {\bf 365}, 407 (1993).
\bibitem{safar95} H. Safar et al., Phys. Rev. B {\bf 52}, 6211 (1995).
\bibitem{ryu96} S. Ryu, A. Kapitulnik and S. Doniach,
  Phys. Rev. Lett. {\bf 77}, 2300 (1996).
\bibitem{safar93} H. Safar et al.,
  Phys. Rev. Lett. {\bf 70}, 3800 (1993).
\bibitem{kwok94} W.~K. Kwok et al.,
  Phys. Rev. Lett. {\bf 73}, 2614 (1994).
\bibitem{jagla96} E. Jagla and C. Balseiro,
  Phys. Rev. Lett. {\bf 77}, 1588 (1996).
\bibitem{wilkin97} N. Wilkin and H. Jensen,
  Phys. Rev. Lett. {\bf 79}, 4254 (1997).
\bibitem{young97} C. Wengel and A. Young,
  Phys. Rev. B {\bf 56}, 5918 (1997).
\bibitem{giam2} T. Giamarchi and P. Le Doussal,
  Phys. Rev. B {\bf 55}, 6577 
  (1997); A. Koshelev and V. Vinokur, cond-mat/9801144.
\bibitem{future} A. van Otterlo, R.~T. Scalettar,
  and G.~T. Zim\'{a}nyi, unpublished.
\bibitem{koch} R. Koch et al., Phys. Rev. Lett. {\bf 63}, 1511 (1989).
\bibitem{lopez98} D. Lopez et al.,
  Phys. Rev. Lett. {\bf 80}, 1070 (1998).
\end{references}
\end{document}